\documentclass[aps,prd,twocolumn]{revtex4-1}

\usepackage{aas_macros}
\usepackage{epsfig}
\usepackage{latexsym}
\usepackage{times}

\newcommand{\bed}{\begin{displaymath}}
\newcommand{\eed}{\end{displaymath}}
\newcommand{\bef}{\begin{figure}}
\newcommand{\eef}{\end{figure}}
\newcommand{\ben}{\begin{enumerate}}
\newcommand{\bei}{\begin{itemize}}
\newcommand{\eei}{\end{itemize}}
\newcommand{\een}{\end{enumerate}}
\newcommand{\beq}{\begin{equation}}
\newcommand{\eeq}{\end{equation}}
\newcommand{\ber}{\begin{eqnarray}}
\newcommand{\eer}{\end{eqnarray}}

\newcommand{\gcc}{\mbox{${\rm gm.} \, {\rm cm}^{-3}$}}
\newcommand{\mdot}{\mbox{$\dot{\rm M}$}}
\newcommand{\msun}{\mbox{{\rm M}$_{\odot}$}}

\newcommand{\lsim}{\raisebox{-0.3ex}{\mbox{$\stackrel{<}{_\sim} \,$}}}
\newcommand{\gsim}{\raisebox{-0.3ex}{\mbox{$\stackrel{>}{_\sim} \,$}}}

\usepackage{xcolor}
\newcounter{attnctr} 

\usepackage{ulem}

\begin{document}

\title{Of Mountains and Molehills :
      {\sl Gravitational Waves from Neutron Stars}}
\author{Sushan Konar}
\affiliation{NCRA-TIFR, Pune, India} 
\email[]{sushan@ncra.tifr.res.in}
\homepage[]{http://www.ncra.tifr.res.in:8081/~sushan/}
\author{Dipanjan Mukherjee}
\email[]{dipanjan.mukherjee@anu.edu.au}
\affiliation{RSAA-ANU, Canberra, Australia}
\author{Dipankar Bhattacharya}
\email[]{dipankar@iucaa.in}
\affiliation{IUCAA, Pune, India}
\author{Prakash Sarkar}
\email[]{prakash.sarkar@gmail.com}
\affiliation{NIT, Jamshedpur, India}

\begin{abstract}
Surface asymmetries  of accreting  neutron stars are  investigated for
their mass  quadrupole moment  content.  Though  the amplitude  of the
gravitational waves from such asymmetries  seem to be beyond the limit
of detectability  of the present  generation of detectors,  it appears
that  rapidly  rotating  neutron  stars with  strong  magnetic  fields
residing in HMXBs  would be worth considering for  targeted search for
continuous   gravitational  waves   with   the   next  generation   of
instruments.
\end{abstract}

\pacs{}


\maketitle

\date{\today}

\section{introduction}

Definitive   detection   of   gravitational   waves   from   colliding
stellar-mass  black holes~\cite{abbot16b,abbot16c}  have ushered  in a
new  era  of astronomy  and  astrophysics  by  opening up  a  hitherto
unexplored waveband. It is but expected that the interest in this area
would escalate in the coming decades with plans for even more advanced
detectors.    However,   these   collision  events   are   short-lived
transients. Targeted  observations of steady sources  of gravitational
waves are  still keenly awaited.   As those would allow  for excellent
opportunities to  study both the  emission processes and  the emitting
objects in great detail.

Neutron stars  are hypothesised to  be prolific and steady  sources of
gravitational  waves   (see  \cite{lasky15}   for  a   brief  review),
particularly  because  of  their   extreme  compactness  and  enormous
magnetic  fields.  Gravitational  waves  were  originally invoked  for
neutron stars residing  in low-mass X-ray binaries  (LMXBs) to explain
the  absence of  neutron stars  with spin  frequencies close  to their
break-up  limit  of   $\sim  10^3$~Hz~\cite{bilds98,chakr03}.   It  is
understood that some  of the bright neutron stars  accreting closer to
the Eddington limit  could possibly be detected by  the advanced Laser
Interferometer  Gravitational-Wave Observatory  (aLIGO)  if they  emit
gravitational   waves   at   rates    that   balance   the   accretion
torques~\cite{watts08a,aasi15}.

Moreover  non-axisymmetric  neutron  stars are  expected  to  generate
continuous  gravitational  waves   as  almost  monochromatic  signals.
Evidently  these  would make  for  excellent  candidates for  targeted
gravitational  wave   search  by  advanced  detectors.    This  is  of
importance because direct detection  of such gravitational waves would
impact the  understanding of neutron  star interiors, in terms  of the
equation  of  state  and  material properties  of  dense  matter.   It
therefore  makes  sense to  revisit  some  of the  gravitational  wave
emission scenarios  from non-axisymmetric neutron  stars, particularly
in  view of  some  of the  recent revisions  in  neutron star  crustal
physics~\cite{horow09b,chugu10}.

Structural asymmetries  of neutron stars  have been of interest  for a
long time and  have received attention in different  contexts like the
evolution                          of                         magnetic
fields~\cite{brown98a,melat01,choud02,konar04b,payne04},           the
production      of      kilohertz     quasi-periodic      oscillations
(QPO)~\cite{stuch08}    or    the    generation    of    gravitational
waves~\cite{ushom00,melat05,haske06,vigel10,mastr11,johns13,haske15a,haske15b}.

There  could be  several possible  ways of  creating such  asymmetries
since  the solid  crust of  a neutron  star is  capable of  supporting
deviations  from axisymmetry  by anisotropic  stresses.  The  simplest
situation is  that of  the existence of  surface mountains  similar to
those seen on solid planets~\cite{scheu81}. Strong magnetic fields can
also distort  the star if  the magnetic axis  is not aligned  with the
axis of rotation.  Indeed the  effects of strong higher multi-poles and
toroidal  components   of  the  magnetic  field   in  generating  such
non-axisymmetries            have             recently            been
investigated~\cite{lasky13,mastr13,mastr15}.   Other   mechanisms  for
generating   asymmetries  could   be   the   development  of   dynamic
instabilities  in rapidly  rotating  neutron stars  driven by  nuclear
matter  viscosity~\cite{bonaz96} or  through r-mode  oscillations (see
\citep{haske15b} and references therein).

Conceivably any  of the above  mechanisms can  be active in  a neutron
star. Accreting neutron  stars can have yet another  additional set of
mechanisms for producing non-axisymmetry.   It has been suggested that
non-axisymmetric temperature  variations in the crust  of an accreting
neutron star could lead to  `wavy' electron capture layers giving rise
to    horizontal     density    variations    near     such    capture
layers~\cite{bilds98,ushom00}.

However, the most  widely discussed scenario for  an accreting neutron
star to have  a non-zero quadrupole moment (leading  to the generation
of  gravitational waves)  is  that of  the  accretion induced  surface
mountains         supported         by         strong         magnetic
fields~\cite{brown98a,melat01,payne04,vigel08}.   Usually,  ultra-fast
($P_s \sim $ms) neutron stars in  LMXBs are considered in this context
because such  short spin  periods (achievable only  in LMXBs)  imply a
larger  amplitude   of  the   emitted  gravitational   waves.   Recent
investigations  however  indicate  that   the  prospect  of  detecting
gravitational  waves from  LMXBs is  not very  encouraging unless  the
neutrons  star has  a buried  magnetic  field of  $\sim 10^{12}$~G  or
more~\cite{haske15a},  even though  the existence  of buried  magnetic
fields  of such  strength is  thought to  be unlikely  in presence  of
magnetic buoyancy~\cite{konar04b}.   On the other hand,  neutron stars
in high  mass X-ray binaries  (HMXB) typically have  stronger magnetic
fields than  in LMXBs  and therefore  accretion induced  mountains are
likely to  be larger in  them, giving  rise to larger  mass quadrupole
moments.   However the  neutron  star  spin-frequencies are  typically
observed to be in the  range 1--10$^{-3}$~Hz in HMXBs.  Therefore, any
gravitational  wave arising  from  structural  asymmetries in  neutron
stars residing  in HMXBs  are likely  to be detected  only by  the new
generation of detectors accessing such a frequency range.

In a series of  recent papers~\cite{mukhe12,mukhe13a,mukhe13b}, two of
the present authors have investigated  the nature of accretion induced
mountains  on  neutron  stars  in  HMXBs.  We  use  these  results  to
investigate  the gravitational  wave  emission from  neutron stars  in
HMXBs   due   to   the  magnetically   confined   accretion   columns.
Accordingly, this paper is organised as follows. In Sec.\ref{sec_mntn}
we revisit the  crustal mountains on the surface of  the neutron stars
in  view   of  the   recent  revisions   in  crustal   properties.  In 
Sec.\ref{sec_accn}  magnetically   confined  mountains   in  accreting
neutron stars  are investigated.   We estimate  the amplitudes  of the
gravitational waves that  could be produced by such  structures on the
surface   and  consider   the  possibility   of  their   detection  in
Sec.\ref{sec_gwve}   Finally  our   conclusions   are  summarised   in
Sec.\ref{sec_cncl}.

\section{surface asymmetries}

\subsection{A Crystal Mountain}
\label{sec_mntn}

The  crust of  a  neutron star  is essentially  solid,  apart from  an
extremely thin liquid  surface layer. Consequently, its  shape may not
be  necessarily axisymmetric  as  deviations from  axisymmetry can  be
supported  by anisotropic  stresses in  the solid.   The shape  of the
crust depends  not only  on the  geological history  of the  star (for
example, episodes  of crystallisation) but  also on star quakes.  As a
result, crystalline  mountains may exist  on the surface of  a neutron
star, similar to those on the surface of a solid planet.

It has  been demonstrated~\citep{scheu81}  that in a  homogeneous rock
stable mountains  can not rise  much further  than $h_1 =  Y/\rho g_s$
above the  level of the  surrounding plains  ($\rho$ - density  at the
base of the  mountain, $g_s$ - surface gravity, $Y$  - yield stress of
the crustal material).  Gently sloping hills of crustal rock, floating
in more or  less isostatic conditions on denser material,  may be able
to rise to greater heights of the order of $h_2 = (h_1 b)^{1/2}$ where
$b$ is the  width of the base.  From these  considerations the maximum
height of such a mountain, on the  surface of a neutron star, had been
estimated  to be  $  \sim 10^{-3}$~cm.   However, certain  assumptions
about  the properties  of the  crustal material  are inherent  in this
estimate and  these assumptions  require another look  considering the
recent revisions of those properties.

Assuming the  basic nature of a  mountain on the surface  of a neutron
star  to be  the same  as a  mountain on  a rocky  planet we  take the
average density of the mountain material to be the same as that of the
base,  i.e.,  the surface  density  of  the  star. The  condition  for
stability of such  a mountain is that the pressure  at the base, given
by
\ber
P_{\rm m} = \rho_{\rm m} \, g_{\rm ns} \, h_{\rm m},
\label{eq_pmnt}
\eer
is less  than the shear stress  of the material on  the surface; where
$\rho_{\rm m}$ is the average density of the mountain, $g_{\rm ns}$ is
the surface gravity of  the star and $h_{\rm m}$ is  the height of the
mountain. We  assume the  base of  the mountain to  be located  at the
outermost solid surface layer, where the liquid-solid phase transition
occurs.  This phase-transition is expected  to take place when $\Gamma
\simeq 175$,  where $\Gamma \,  (= Z^2  e^2 / a  k_{\rm B} T)$  is the
Coulomb coupling parameter, $Z$ is  the dominant ionic species at that
density  and $a$  is the  lattice  spacing of  the solid.   Typically,
$\rho_m  \sim  10^7  -  10^8$~\gcc for  a  cold  (surface  temperature
$10^6$~K)   neutron    star.    The    surface   gravity    is   $\sim
10^{14}$~cm.s$^{-2}$  for a  typical  neutron star  of mass  1.4~\msun
\ and radius 10~Km.

The `yield' or shear stress of the  material in the crust of a neutron
star is given by,
\beq
S = \mu \theta \,,
\eeq
where $\mu$ is  the shear modulus and $\theta$ is  the shear strain of
the crust thought to be made up of unscreened nuclei arranged in a bcc
`metallic'  lattice in  which the  inter-nuclear spacing  (varies from
$10^{-9} -  10^{-11}$~cm in the  density range $10^5  - 10^{10}$~\gcc)
exceeds the  nuclear size  by several orders  of magnitude.   Thus the
lattice is  very `open' and `Coulombic'  in nature.  It has  also been
argued  that  the crust  of  a  neutron star  may  exist  in a  glassy
state~\cite{stroh91}  but the  effective shear  modulus averaged  over
directions  are not  very  different for  the bcc  crystal  and for  a
quenched glassy solid. (Though  recent observations of cooling neutron
star transients imply that the crust  of a neutron star is unlikely to
be in an amorphous glassy state~\cite{brown09,page13}.)

The shear modulus of this Coulomb crystal is given by~\cite{stroh91},
\ber
\mu = \frac{0.1194}{1+ 1.781\times (100/\Gamma)^2} \frac{n(Ze)^2}{a},
\eer
where, $n$ is the ion number  density, $a$ is the inter-ionic distance
(lattice  spacing), $Z$  is the  atomic weight  of the  dominant ionic
species and  $\Gamma$ is  the Coulomb  coupling parameter.   Since the
dominant  ionic species  at a  density  of $10^8$~\gcc  happens to  be
$_{28}Ni^{62}$~\cite{baym71c} (non-accreted crust),  we typically have
:
$Z = 28, \; \; n \simeq 10^{30}~{\rm cm}^{-3} 
        \; \; \mbox{and} \; \; a \simeq 10^{-10}~{\rm cm}$.

Even  though  the  calculation  of the  shear  modulus  is  relatively
straightforward, obtaining  the correct value  of the shear  strain is
somewhat  complicated.  Theoretical  calculations based  on chemically
and  crystallographically  perfect  crystals of  `terrestrial'  metals
gives $\theta \simeq 10^{-1} - 10^{-2}$~\cite{smolu70a}.  But $\theta$
is  a   `structurally  sensitive'  quantity  and   its  value  changes
significantly  from an  ideal crystal  to that  containing impurities,
defects etc. The  material in the pulsar crust  has traditionally been
considered  to be  chemically impure  (result of  partial burning  and
incomplete mixing).  It is also expected  to contain a large number of
defects such  as grain  boundaries, dislocations  etc.  which  may get
enhanced  in older  stars  that  have experienced  a  large number  of
star-quakes  and/or  glitches.   These defects/dislocations  are  also
expected to  lower the value of  $\theta$ and the shear  strain on the
surface of the neutron star has been estimated to be \cite{smolu70b},
\ber
\theta \sim 10^{-4} - 10^{-3} \,,
\eer
the upper limit coming from the  glitch magnitudes of radio pulsars in
conformity with the star-quake hypothesis~\cite{ruder91b}.

Then the shear  stress on the surface of a  typical neutron star turns
out to be,
\beq
S  = \mu  \theta  \simeq  2 \times  10^{20}  -  2 \times  10^{21}~{\rm
  dyne.cm}^{-2} \,.
\eeq
The maximum height a mountain on the surface of a neutron star is then
obtained when $P_{\rm mnt} \simeq S$, and is given by,
\beq
h_{\rm mnt}^{\rm max} \simeq 0.02 - 0.2~{\rm cm} \,.
\eeq
The maximum mass contained in such a mountain would be,
\beq
\Delta M
\sim \rho_{\rm mnt} (h^{\rm max}_{\rm mnt})^3
\sim 10^8~{\rm gm} \sim 10^{-25} M_{\rm ns} \,,
\eeq
which is tiny compared to the total stellar mass.

As mentioned  above this  estimate assumed the  neutron star  crust to
harbour a  large number of defects  and/or dislocations and to  have a
high         impurity         content.         However,         recent
investigations~\cite{horow09b,chugu10},   that   have  made   use   of
molecular  dynamic simulations,  indicate  that due  to the  extremely
high-pressure  environment  the  crust  would mostly  be  in  a  `pure
crystal'  phase. Accordingly,  the breaking  strain (maximum  value of
shear  strain, $\theta$,  that can  be withstood  by the  material) is
found to be $\sim 0.1$, much  larger than the previous estimates. This
implies that  the maximum  height of  a mountain on  the surface  of a
neutron star is  $\sim 2 - 20$~cm,  taking the mass content  of such a
mountain to,
\beq
\Delta  M   
\sim  \rho_{\rm   mnt}  (h^{\rm  max}_{\rm   mnt})^3  
\sim 10^{14}~{\rm gm} \sim 10^{-19} M_{\rm ns} \,.
\eeq
Though this is  much larger than the previous  estimate, the resulting
mass  quadrupole moment  is still  about $10^{33}$~gm.cm$^{2}$  giving
rise to gravitational  waves that would cause a strain  of around
$\sim 10^{-42}$ in a detector (see Sec.\ref{sec_gwve} for details of
strains caused by gravitational waves etc.).

It needs  to be noted  here that all of  the above estimates  are made
assuming an isolated, cold neutron  star with a surface temperature of
$\sim 10^6$~K.  This is  appropriate because  the breaking  strain has
been shown to be highly  dependent on temperature and thermal history,
but  to  remain  more  or less  constant  at  $10^6$~K~\cite{chugu10}.
Evidently,  these  estimates would  not  hold  good for  an  accreting
neutron  star  where the  surface  temperature  could  be as  high  as
$10^8$~K,  at which  temperature the  breaking strain  is expected  to
change  by  half  an  order  of magnitude  over  the  timescale  of  a
year.  Even  though the  solid  surface  layer  would move  to  higher
densities in  a hotter star, changing  the naive estimates above  by a
few orders  of magnitudes, any  stable mountain  on the surface  of an
accreting neutron star may not necessarily be much larger owing to the
variation in the breaking strain.

It is evident  that crystalline mountains, or  rather {\it molehills},
on the crust of neutron stars have ridiculously small mass content and
hence  are   totally  unsuitable   for  any   gravitational  radiation
experiment.  Fortunately,  other effects become important  in creating
larger mountains in accreting neutron stars.  We discuss the situation
in the next section.

\subsection{Magnetically confined Accretion Columns}
\label{sec_accn}

Early on, it has been shown \cite{woosl82, hameu83} that the accretion
column on a neutron star is  like a small mountain of ionised hydrogen
over  the  polar  cap,  supported  by a  strong  magnetic  field.   In
particular, in HMXBs the accreting  material passes through a shock to
finally settle onto the polar cap.  This may or may not happen at both
the  poles.  But  the height  of such  a column  is restricted  by the
condition that material  starts flowing sideways when  the pressure in
the  accretion column  becomes large  enough (typically  hundred times
more  than  the  magnetic  pressure  responsible  for  supporting  the
column~\cite{brown98a}) to bend the  magnetic field lines sufficiently
producing strong horizontal components.

Let us consider a neutron star with a magnetic field strong enough for
the accretion  to be polar  (typically $B \gsim 10^{11}$~G  in HMXBs).
The material flows  in along open field lines and  reaches the surface
within the  polar cap region.  If  we further assume the  rotation and
the magnetic  axis to  be aligned  then tracing  the footprint  of the
dipolar  field lines  onto the  stellar surface,  from where  material
(accretion disk)  stresses and  magnetic stresses are  in equilibrium,
the area of  the polar cap, $A_P$,  of a neutron star  of mass $M_{\rm
  ns}$     and    radius     $R_{\rm    ns}$     is    obtained     to
be~\cite{shapi83,konar97c},
\ber  
A_P  &=&  \pi  R_{\rm ns}^3 R_A^{-1}  \nonumber  \\  
     &=&  \pi (2G)^{1/7} M_{\rm ns}^{1/7} R_{\rm ns}^{9/7} B_s^{-4/7} {\mdot}^{2/7} \nonumber \\
     &=& 2.17 \times 10^{11}
         \left(\frac{B_s}{10^{12}~{\rm G}}\right)^{-4/7} 
         \left(\frac{\mdot}{10^{-8}~\msun/{\rm yr}}\right)^{2/7} \nonumber \\
    &&   \times \left(\frac{M_{\rm ns}}{1.4~\msun}\right)^{1/7} 
         \left(\frac{R_{\rm ns}}{10^6~{\rm cm}}\right)^{9/7} \, {\rm cm}^2 \,,
\eer
where $B_s$  is the strength of  the surface field, $\mdot$ is the
rate of accretion and $R_A$ is the Alfven radius given by
\beq
R_A = (2GM_{\rm ns})^{-1/7} R_{\rm ns}^{12/7} B_s^{4/7} {\mdot}^{-2/7} \, {\rm cm} \,.
\eeq

We determine the extent of the accretion column by
the  condition that  the  pressure  at the  bottom  of  the column  is
$\sim100$ times as large as the magnetic pressure~\cite{brown98a}, i.e,
\ber
P_{\rm ac}
\sim 4 \times 10^{24} 
\left(\frac{B_{\rm s}}{10^{12}{\rm G}}\right)^2 \, {\rm dyne.cm}^{-2} \,.
\eer
where $P_{\rm ac}$ is the pressure at the bottom of the column.  
\bef
\epsfig{file=bps.ps,width=150pt,angle=-90}
\caption[]{The pressure vs.  density in the outer layers  of a neutron
  star, as calculated by \citet{baym71c}.  The stars correspond to the
  actual  EoS, whereas  the  solid curve  corresponds  to our  fitting
  formula given by Eq.[\ref{eq_bps}].}
\label{fig_bps}
\eef
Hydrostatic equilibrium demands that this pressure equals the pressure
in the crust  elsewhere (outside the polar cap) at  the same radius as
the bottom  of the column.  This  pressure is due to  the relativistic
degenerate electrons and  the ions, given by  the appropriate equation
of state~\cite{baym71c}.   Interestingly, the pressure in  the density
range  $10^6~\gcc  \leq   \rho  \leq  10^9~\gcc$  can   be  very  well
approximated by the following fitting formula,
\ber
\log P &=& 13.65 + 1.45 \log \rho \,.
%
\label{eq_bps}
\eer
As  can  be  seen  from  Fig.[\ref{fig_bps}],  this  is  a  reasonable
approximation.  This is significant.  Because, for the entire range of
accretion  rate   realisable  in   a  neutron  star   ($10^{-14}  \leq
\mdot/\msun.{\rm yr}^{-1}  \leq 10^{-8}$) the density  above which the
crustal material remains solid (and  consequently the density at which
the accretion column  would be anchored) is confined to  this range of
densities~\cite{konar97c}.  Therefore, the  relation between the field
strength and  the density  at the  bottom of  the accretion  column is
roughly given by,
\ber
%
\left(\frac{\rho}{10^7\gcc}\right)
&=& 6.34 \, \left(\frac{B_{\rm s}}{10^{12}{\rm G}}\right)^{1.38} \,.
\eer
The scale height of the column, $h_{\rm ac}$, then turns out to be,
\ber
h_{\rm ac}
&=& \frac{P_{\rm ac}}{\rho_{\rm bot} \, g_{\rm ns}} \nonumber \\
&=& 3.4 \times 10^2 {\rm cm} \, \nonumber \\
&&  \times \left(\frac{B_{\rm s}}{10^{12}~{\rm G}}\right)^{0.62}
    \left(\frac{M_{\rm ns}}{1.4~\msun}\right)^{-1}
    \left(\frac{R_{\rm ns}}{10~{\rm km}}\right)^2 .
\eer
If we describe  the density profile within the accretion  column by an
`atmosphere'  solution~\cite{bilds95}  then the  total mass  contained
within the column is given by,
\ber
M_{\rm ac} 
&=& A_P \int_h^0 \rho_{\rm bot} \, e^{-x/h} dx  \nonumber \\
&=& 2.95 \times 10^{21} {\rm gm} \, 
    \left(\frac{M_{\rm ns}}{1.4~\msun}\right)^{-6/7} 
    \left(\frac{R_{\rm ns}}{10~{\rm km}}\right)^{23/7} \nonumber \\
    && \times \left(\frac{B_s}{10^{12}~G}\right)^{10/7} 
    \left(\frac{\mdot}{10^{-8}~\msun/{\rm yr}}\right)^{2/7} \nonumber \\
&=& 1.5 \times 10^{-12} \, \msun
    \left(\frac{M_{\rm ns}}{1.4~\msun}\right)^{-6/7} 
    \left(\frac{R_{\rm ns}}{10~{\rm km}}\right)^{23/7} \nonumber \\ 
    && \times \left(\frac{B_s}{10^{12}~G}\right)^{10/7} 
    \left(\frac{\mdot}{10^{-8}~\msun/{\rm yr}}\right)^{2/7} .
\label{eq_mass}
\eer

It    needs    to    be    noted   that    some    of    the    recent
estimates~\cite{payne04,payne07,vigel08}   place   the   mass   of   a
magnetically confined accretion column at a much larger value ($\Delta
M  \sim  10^{-5}$~\msun)  by  allowing  for  mass-loading  beyond  the
accretion column.   This approach makes  use of plasma loading  on all
field  lines  providing  additional   lateral  support  to  help  form
accretion mounds  with very  large masses.  However  MHD instabilities
~\cite{litwi01,cummi01}  are expected  to play  a significant  role in
determining the  extent of  the accretion  column, since  the accreted
material is  expected to flow  towards the equator not  over accretion
time-scales (which  could be  very large) but  over much  smaller flow
time-scales. An early and rough estimate placed the value of this flow
time-scale to about a year~\cite{choud02,konar04b}. On the other hand,
this flow time-scale happens to  be larger than the dynamic time-scale
of the neutron  star -- which simply means that  the accreted material
would be  assimilated as fast or  faster than it comes  to the equator
and  the density  of  the crust  would adjust  itself  to a  spherical
profile  before  it  becomes  asymmetric  enough to  give  rise  to  a
significant quadrupole moment.

It should be  mentioned here that our use of  an `atmospheric' density
profile inside  the accretion column inherently  assumes an isothermal
situation.  However,  it has  been shown that  the correct  physics is
obtained by assuming an adiabatic process.  By numerically solving the
appropriate Grad-Shafranov equation it has also been demonstrated that
magneto-static solutions cannot be  found for accretion columns beyond
a  threshold height  (and  mass)  indicative of  the  presence of  MHD
instabilities~\cite{mukhe12}.   The accreted  matter would  eventually
flow horizontally  along the neutron  star surface.  Detailed  two and
three-dimensional MHD  simulations~\cite{mukhe13a,mukhe13b} indicate a
lower  mass threshold  ($5  \times 10^{-13}$~\msun  \/  for $B_p  \sim
10^{12}$~G)  above which  pressure  driven  instabilities would  start
operating and  matter could not  be efficiently confined by  the local
field in the polar cap.  This  threshold mass is much smaller than the
amount               indicated                in               earlier
investigations~\cite{payne04,payne07,vigel08},   but    matches   very
closely  with  that  obtained  from simple  dimensional  estimates  in
Eq.[\ref{eq_mass}].   This allows  us to  use Eq.[\ref{eq_mass}],  for
obtaining approximate  values of the  mass of an accretion  column, in
the rest of this investigation.

\section{gravitational waves}
\label{sec_gwve}

The  amplitude of  a gravitational  wave is  described in  terms of  a
strain, a dimensionless quantity $h$. This gives a fractional change in
length, or  equivalently light  travel time,  across a  detector.  The
maximum  amplitude  of  gravitational  waves produced  by  a  spinning
neutron star  due to a  structural asymmetry  in the weak  field, wave
zone limit ($d \, \gsim c P_s$) is given by~\cite{bonaz96},
\ber
h_0 = 24 \pi^2 G  c^{-4} P_s^{-2} {\cal Q} \, d^{-1} \,,
\eer
where $P_s$ is the spin-period of the star, $d$ is the distance of the
observer from  the star and ${\cal  Q}$ is the mass  quadrupole moment
about the principal axis of asymmetry.   It should be noted that $\cal
Q$ is the  stationary mass quadrupole moment in the  rotating frame of
the  star and  leads to  the emission  of gravitational  waves with  a
predominant frequency $\nu = 2/P_s$~\cite{shapi83}.

\begin{table}
\begin{tabular}{|c|c|c|c|c|} \hline
$B_s$  & $z_c$  &  $M$  &  $\epsilon$ &  $h_0$  \\
G      &  cm    & \msun &             & \\
&&&&\\  
$1 \times 10^{11}$ & $15$ & $5.67 \times 10^{-14}$ & $1.12 \times 10^{-13}$ & $4.71 \times 10^{-37}$ \\ 
$5 \times 10^{11}$ & $30$ & $4.63 \times 10^{-13}$ & $9.20 \times 10^{-13}$ & $3.87 \times 10^{-36}$ \\ 
$1 \times 10^{12}$ & $43$ & $1.52 \times 10^{-12}$ & $3.02 \times 10^{-12}$ & $1.27 \times 10^{-35}$ \\ 
$5 \times 10^{12}$ & $72$ & $8.13 \times 10^{-12}$ & $1.62 \times 10^{-11}$ & $6.82 \times 10^{-35}$ \\ \hline
\end{tabular}
\caption[]{Expected  amplitude  of  gravitational  waves  due  to  the
  magnetically  confined  accretion  columns   on  a  typical  neutron
  star. The  quantities in  the columns ($B_s,  z_c, M,  \epsilon, h_0$)
  respectively refer to - a) the surface magnetic field, b) the height
  of the accretion  column, c) the mass content of  the column, d) the
  ellipticity of the  star due to the existence of  the column, and e)
  the amplitude of the gravitation waves. The spin-period, $P_s$ and the
  distance of the star, $d$ have respectively been assumed to be 1~s and
  1~Kpc for all of the above cases.}
\label{tab_eps}
\end{table}

The mass quadrupole  moment is defined as the quadrupolar  part of the
$1/r^3$ term of the $1/r$ expansion of the metric coefficient $g_{00}$
in  an asymptotically  Cartesian and  mass centred  coordinate system.
The quadrupole  moment of the magnetically  confined accretion column,
considered in Sec.\ref{sec_accn}, is estimated to be,
\ber
{\cal Q} 
&=& M_{\rm ac} R_{\rm ns}^2 \nonumber \\
&=& 2.96 \times 10^{33} {\rm gm.cm}^2 \,
    \left(\frac{M_{\rm ns}}{1.4~\msun}\right)^{-6/7}
    \left(\frac{R_{\rm ns}}{10~{\rm km}}\right)^{23/7+2} \nonumber \\
    && \times \left(\frac{B_s}{10^{12}~G}\right)^{10/7} 
    \left(\frac{\mdot}{10^{-8}~\msun/{\rm yr}}\right)^{2/7} .
%
\eer
This implies that the amplitude of the gravitational wave generated by
such accretion columns would be,
\ber
h_0
&=& 1.87 \times 10^{-35}
    \left(\frac{M_{\rm ns}}{1.4~\msun}\right)^{-6/7}
    \left(\frac{R_{\rm ns}}{10~{\rm km}}\right)^{37/7}   \nonumber \\
    && \times \, P_s^{-2} \left(\frac{B_s}{10^{12}~G}\right)^{10/7}  \nonumber \\
    && \times \left(\frac{\mdot}{10^{-8}~\msun/{\rm yr}}\right)^{2/7} 
    \left(\frac{d}{\rm kpc}\right)^{-1} \,.
%
\label{eq_hpb}
\eer
Now, the wave  amplitude is usually  expressed in terms of  the ellipticity
($\epsilon$) of the non-axisymmetric star, defined as
\beq
I \epsilon = \frac{3}{2} {\mathcal{Q}}_{zz},
\eeq
where $I$ is the moment of inertia of the star. Then the wave amplitude
can be written as,
\ber
h_0
&=& \frac{16\pi^2G}{c^4} \frac{I\epsilon}{P_s^2d} \nonumber \\
&=& 4.21 \times 10^{-30}
   \left(\frac{I}{10^{45}~{\rm gm.cm}^2}\right) 
   \left(\frac{\epsilon}{10^{-6}}\right) \nonumber \\
   && \times \, P_s^{-2} \left(\frac{d}{\rm kpc}\right)^{-1} \,.
\eer
Assuming the accretion column to peak in the $z$-direction, the 
asymmetry $\epsilon$ is given by,
\ber
\epsilon
&=& \left|\frac{I_{zz} - I_{xx}}{I_{xx}}\right|
 = \left|\frac{\Delta I_{zz} - \Delta I_{xx}}{I_0+ \Delta I_{xx}}\right| \,,
\eer
where,  $I_{aa}$   is  the   principal  moment   of  inertia   in  the
$a$-direction  and $I_{aa}  =  I_o  + \Delta  I_{aa}$  with $I_o$  and
$\Delta I_{aa}$  being the symmetric  and the asymmetric parts  of it.
Based on the  detailed calculations made in  \cite{mukhe12}, we obtain
precise values of $\epsilon$ and $h_0$ for a few specific cases. These
are shown  in table[\ref{tab_eps}]. It  is seen that the  $h_0$ values
obtained   here  matches   almost  exactly   with  that   obtained  in
Eq.\ref{eq_hpb} which is but an approximation.

\setcounter{figure}{1}
\begin{figure*}
\epsfig{file=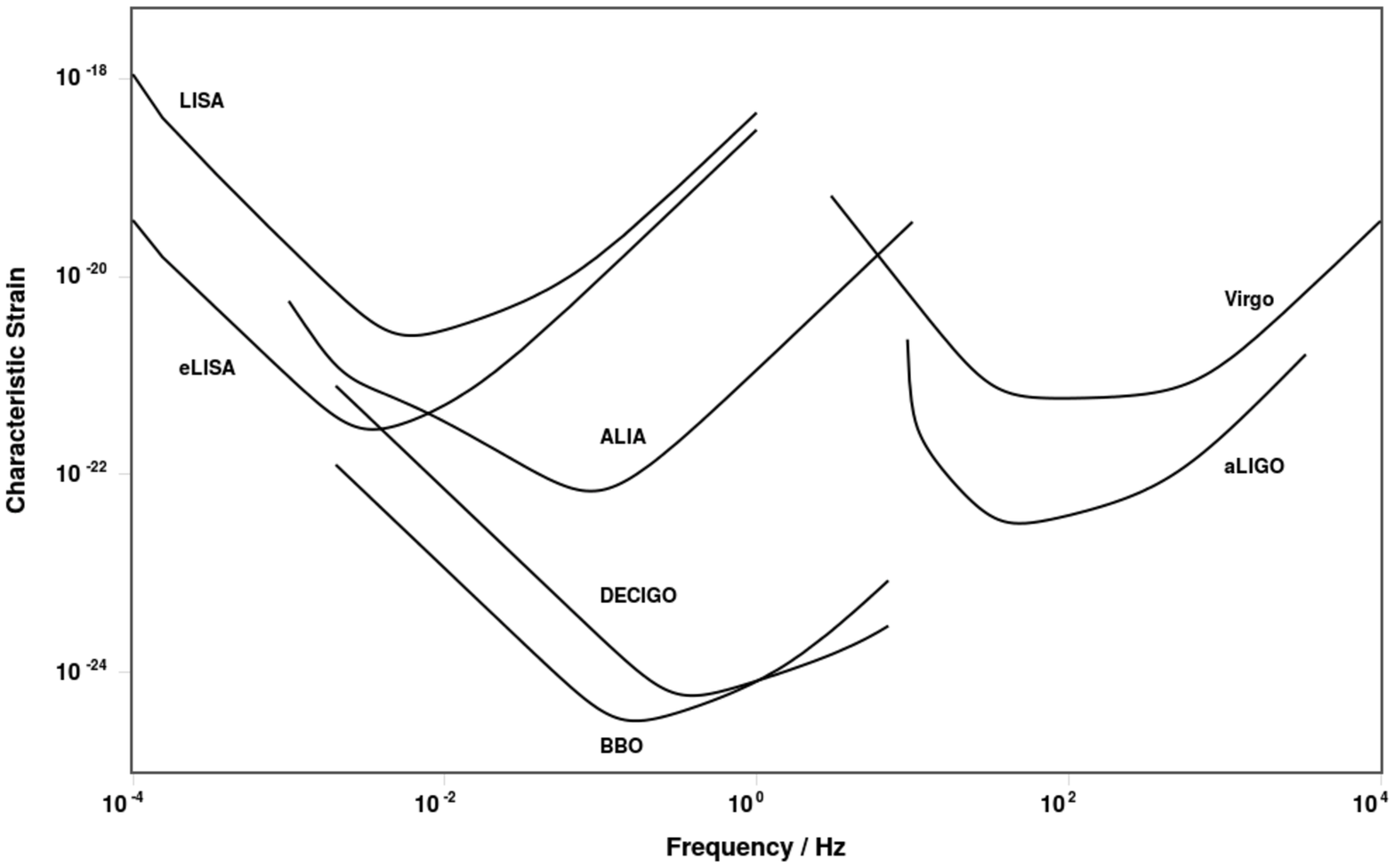,width=450pt}
\caption[]{Sensitivities of  second generation space  and ground-based
  gravitational  wave detectors,  in the  neutron star  spin-frequency
  range.  The  sensitivity curves  have been generated  using publicly
  available  resources at  {\tt  http://www.rhcole.com/apps/GWplotter/}
  \cite{moore15}. }
\label{fig_detector}
\end{figure*}

The spin-periods of neutron  stars (from ultra-fast millisecond pulsars
to slow  X-ray pulsars in HMXBs)  span a range of  $10^{-3} - 10^3$~s.
Hence,  the detectors  appropriate  for  detecting gravitational  wave
signatures   (of    the   kind   discussed   here)    would   be   the
aLIGO~\cite{abadi10}     and     its     extended     version     once
LIGO-India~\cite{unnik13} starts operation. But it can also be readily
seen that  the possibility of  detection would be severely  limited by
the sensitivity of the present day detectors~\cite{abbot09}.  However,
it  can  be  seen  from Fig.[\ref{fig_detector}]  that  the  space-based
detectors,  like -  the  evolving Laser  Interferometer Space  Antenna
(eLISA)~\cite{elisa13}, Advanced  Laser Interferometer  Antenna (ALIA),
Big Bang  Observer (BBO) and Deci-hertz  Interferometer GW Observatory
(DECIGO)~\cite{moore15}, would  be very good candidates  for this kind
of work  as these would have  much higher sensitivities in  a range of
frequencies that is of interest in the present context. In particular,
the frequency range relevant for neutron stars residing in HMXBs ($\nu
\sim 10^{-3} - 1$~Hz) is precisely  the one in which these space-based
detectors    would   be    operative,    as   can    be   seen    from
Fig.[\ref{fig_detector}] and Fig.[\ref{fig_bp}].

It is  then interesting to find  the parameter space (in  terms of the
spin and the magnetic field of  the neutron stars) that is most likely
to be sampled by future detectors. Let us assume the neutron stars, in
consideration, to have : $M_{\rm ns} = 1.4~\msun, R_{\rm ns} = 10~{\rm
  Km}$,  accreting  at  the  Eddington  rate  and  at  a  distance  of
1~Kpc. Then from Eq.[\ref{eq_hpb}] we obtain,
\beq
h_0 = 1.87 \times 10^{-35} P_s^{-2} \left(\frac{B_s}{10^{12}~G}\right)^{10/7}.
\label{eq_hpb1}
\eeq
This gives rise to a relation between $P_s$ and $B_s$ for a given value
of $h_0$ with the following form,
\beq
\log_{10} B_s = 1.4 \log_{10} P_s + 0.7 \log_{10} h_0 + 36.31 \,,
\label{eq_detect}
\eeq
enabling us to  identify the possible region, in  the $P_s-B_s$ plane,
where future searches could focus on.

\setcounter{figure}{2}
\begin{figure*}
\epsfig{file=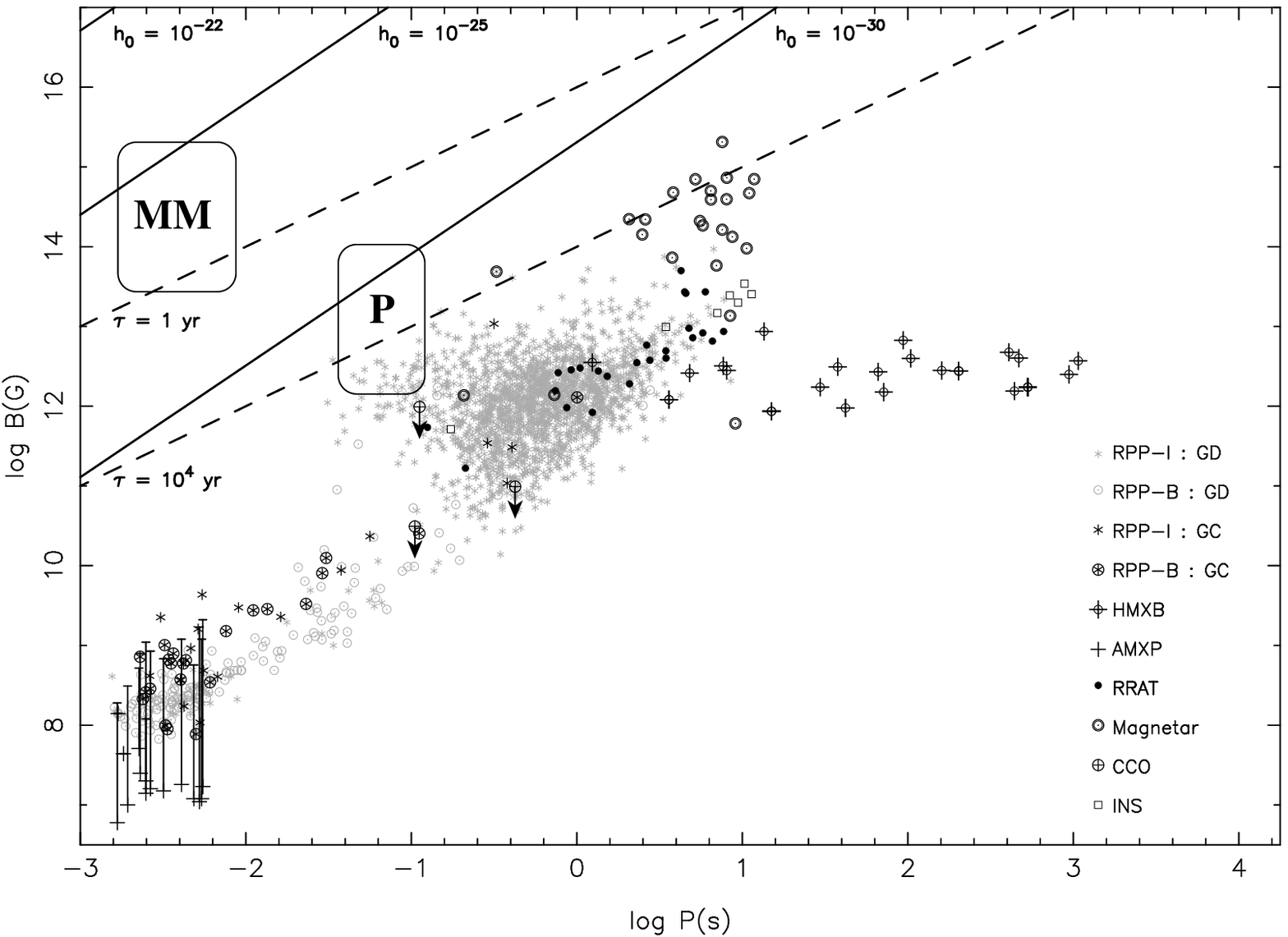,width=450pt}
\caption[]{All    known     neutron    stars    (for     which    some
  measurement/estimate of the magnetic  field exists) in the $P_s-B_s$
  plane  (a detailed  discussion on  various observational  classes of
  neutron stars  can be  found in  \cite{konar16a}).  The  solid lines
  marked `$h_0 = 10^{-22}$', `$h_0 = 10^{-25}$' and `$h_0 = 10^{-30}$'
  are  drawn  using  Eq.[\ref{eq_detect}].    The  dashed  lines  mark
  spin-down  timescales of  1 year  and $10^4$  years.  The  rectangle
  marked  `MM' is  the region  where {\it  millisecond magnetars}  are
  expected to appear, and the  region marked `P' is where non-recycled
  neutron  stars with  strong magnetic  fields and  short spin-periods
  should be seen.   \\ Legends : RPP - rotation  powered pulsar, I/B -
  isolated/binary, GC -  globular cluster, GD - galactic  disc, AMXP -
  accreting millisecond X-ray pulsar (in LMXBs), RRAT - Rotating Radio
  Transients,  INS -  isolated  neutron star,  CCO  - central  compact
  object. Note that  the vertical lines associated with  the AMXPs are
  uncertainties coming  from different  models of field  estimate, not
  error bars. \\
  Data : 
  RPP - \cite{manch05a}, {\tt http://www.atnf.csiro.au/research/pulsar/psrcat/}; \\
  RRAT -  {\tt http://astro.phys.wvu.edu/rratalog/}; \\
  Magnetar - {\tt http://www.physics.mcgill.ca/$\sim$pulsar/magnetar/main.html}; \\
  AMXP - \cite{patru12c,mukhe15}; 
  HMXB - \cite{cabal12}; 
  INS - \cite{haber07,kapla09a}; 
  CCO - \cite{halpe10,ho13}.
}
\label{fig_bp}
\end{figure*}
In Fig.[\ref{fig_bp}]  we plot all  known neutron stars  (accreting or
otherwise)  in  the $P_s-B_s$  plane.   Lines  corresponding to  three
values  of  $h_0$  have  also   been  indicated  in  this  plot  using
Eq.[\ref{eq_detect}].  Since the maximum  sensitivity limits for aLIGO
and  BBO  are  $10^{-22}$   and  $10^{-25}$  respectively,  the  lines
corresponding  to   those  values of  $h_0$ indicate  the maximal
capabilities of  these detectors.  Though  it must be  remembered that
such connection to  a given detector is purely `symbolic'  here as BBO
would  not  even  be  operative  in  the  indicated  frequency  range.
Therefore, the  region to  the left  of a given  line could  simply be
taken to be indicative of the region of possible detectability (by `a'
detector) of  gravitational waves with  amplitudes equal to  or larger
than  the particular  value  of  $h_0$, if  it  is  operative in  that
frequency.   Evidently, no  known neutron  star inhabits  this region.
Moreover, all  the accreting  neutron stars  are very  far away  - the
LMXBs being concentrated in the lower  left hand and the HMXBs showing
up in the upper  middle to upper left hand region  of the neutron star
parameter space.

Looking at Eq.\ref{eq_hpb} or its simplified version Eq.\ref{eq_hpb1},
it is  evident that the  amplitude of the emitted  gravitational waves
increase with  a decrease  in the  spin-period or  an increase  in the
magnetic  field  of  the  neutron  star.   Clearly,  rapidly  rotating
magnetars residing  in HMXBs would  fit the bill perfectly.   It would
also not be remiss to note that to generate the strong magnetic fields
through  a dynamo  process,  the  magnetars are  expected  to be  born
rotating  fast,  with  $P\lsim 1$~ms~\cite{thomp93}.   In  fact,  such
objects  (nicknamed  `millisecond  magnetars') have  been  invoked  to
explain some of the  ultra-luminous soft gamma-ray bursts~\cite{lu14}.
However,   such   objects   have    an   extremely   rapid   rate   of
spin-down. Assuming the spin-down  to be entirely electromagnetic, the
characteristic spin-down timescale is given by~\cite{zhang01a},
\beq
\tau \sim 2 \times 10^3 {\rm s} (I_{45} R_6^{-6}) B_{s, 15}^{-2} P_{s,
  -3}^2,
%
\eeq
where $I_{45}$, $R_6$,  $B_{s, 15}$ and $P_{s, -3}$  denote the moment
of  inertia  in $10^{45}$~gm.cm$^2$,  radius  in  $10^6$~cm, $B_s$  in
$10^{15}$~G    and   $P_s$    in    $10^{-3}$~s   respectively.     In
Fig.[\ref{fig_bp}], two  dashed lines  mark spin-down timescales  of 1
year  and  $10^4$  years  respectively.   Objects on  the  left  of  a
particular line would  have spin-down timescales smaller  than that on
the line itself.  Evidently,  the `millisecond magnetar' phase (region
marked `MM') would be extremely short-lived.

Nevertheless,  strongly magnetised  neutron  stars  residing in  HMXBs
would still be  the best candidates for  gravitational waves generated
due to  the mass quadrupole  moments induced by accretion  columns, as
can be  seen from the proximity  of magnetars to the  $h_0 = 10^{-30}$
line in  Fig.[\ref{fig_bp}] (though this  is orders of  magnitude beyond
the   detection    limit   of   even   the    second   generation   of
detectors). Presence of magnetars in  HMXBs have recently been invoked
to    explain   certain    ultra-luminous   super-giant    fast   X-ray
transients~\cite{bozzo08,musht15}.     Therefore,     such    a
population,  in fact,  does  exist.  Now,  it can  also  be seen  from
Fig.[\ref{fig_bp}]  that the  slow magnetars  have spin-down  timescales
$\sim 10^4$~years  which also happens  to be the typical  timescale of
HMXB activity.

However, it appears that the objects in the region marked `P', neutron
stars with reasonably short $P_s$  and moderately high $B_s$, would be
the best candidates (as sources of steady gravitational waves) if they
can be found in HMXBs  before significant spin-down has happened. Now,
we expect to see an order of magnitude increase in the number of radio
pulsar detection with the advent of the SKA~\cite{smits09,smits11}. It
is conceivable that the region marked  `P' may also see an increase in
the density of  objects with short $P_s$ and high  $B_s$. Detection of
early  HMXB phase  for such  neutron stars  may happen  with the  next
generation of  sensitive X-ray  instruments.  Consequently,  the third
generation of space-based detectors would likely be able to target and
study steady gravitational waves from such systems.

\bef
\epsfig{file=dpol.ps,width=100pt,angle=-90}
\epsfig{file=mpol.ps,width=100pt,angle=-90}
\caption[]{The  left-hand panel  shows a  star-centred dipole  field  in the
range $0 \leq \theta \leq  \pi/2$. The circular arc denotes the extent
of  the  stellar surface.   The  plot shows  the  field  lines up-to  a
distance  of 10  stellar  radius.   In the right-hand panel an  off-axis
(offset by 60$^\circ$) quadrupole field, twice as strong as the dipole
field, has been added.}
\label{fig_mult}
\eef
%


It should  be mentioned  here that  in the  above calculation  we have
assumed a symmetric dipolar magnetic  field.  But a complex field with
very strong higher multipole components near the surface is not ruled
out~\cite{gil01}, as  shown in  Fig.[\ref{fig_mult}].  It  is evident
that a  simple higher multipole  component over and above  the dipole
can  create   other  asymmetric,  off-axis  polar   regions.   Charged
particles moving  along the  field lines  would create  `mountains' at
these positions too. Therefore with a complicated magnetic field it is
possible to obtain a number of  very asymmetric mountains.  And if 
the higher  multipole components  are much  stronger than  the dipole
then the  mass content of the  accretion column could be  much larger.
Recent investigations  have shown that higher  multipoles can generate
large ellipticity and be good candidates for gravitational
waves~\cite{mastr13}.

Moreover, the  estimates in this  section has implicitly  assumed that
the  accretion-induced mountains  would  be stable  even with  surface
magnetic fields much larger than $\sim 10^{13}$~G. This assumption may
not hold if MHD instabilities become  stronger. In that case, the mass
content of the accretion induced mountains and the consequent amplitude
of  the gravitational  waves  would  be even  smaller.  But a  precise
statement in this regard can not be made without detailed calculation
of accretion onto strongly magnetised neutron stars.

\section{conclusion}
\label{sec_cncl}

In this  note we present  simple estimates of mass  quadrupole moments
and corresponding  amplitudes of the  gravitational waves that  can be
generated by a pair of  magnetically confined accretion column.  It is
seen that the wave amplitudes are too small for the present generation
of detectors.   However, rapidly rotating strongly  magnetised neutron
stars in HMXBs are expected to  be good candidates for targeted search
for gravitational waves by the next generation of detectors. \\

\begin{acknowledgments}
The authors would like to  thank Archana Pai for fruitful discussions.
SK is supported by a grant (SR/WOS-A/PM-1038/2014) from the Department
of Science \& Technology, Government of India.
\end{acknowledgments}

\bibliography{adsrefs}

\end{document}